# Lens Absorber Coupled MKIDs for Far Infrared Imaging Spectroscopy

Shahab O. Dabironezare[1,2], Sven van Berkel[3], Pierre M. Echternach[3], Peter K. Day[3], Charles M. Bradford[3], and Jochem J.A. Baselmans[1,2]
[1]Technology group, SRON, Netherlands Institute for Space Research, Leiden, the Netherlands
[2]Terahertz Sensing Group, Microelectronics Dept., Delft University of Technology, Delft, the Netherlands
[3]NASA Jet Propulsion Laboratory (JPL), Pasadena, CA, USA

*Abstract—* Future generation of astronomical imaging spectrometers are targeting the far infrared wavelengths to close the THz astronomy gap. Similar to lens antenna coupled Microwave Kinetic Inductance Detectors (MKIDs), lens absorber coupled MKIDs are a candidate for highly sensitive large format detector arrays. However, the latter is more robust to misalignment and assembly issues at THz frequencies due to its incoherent detection mechanism while requiring a less complex fabrication process. In this work, the performance of such detectors is investigated. The fabrication and sensitivity measurement of several lens absorber coupled MKID array prototypes operating at 6.98 and 12 THz central frequencies is ongoing.

## I. INTRODUCTION

Future generation of cooled space-based observatories are aiming towards hosting Far Infrared (FIR) imaging spectrometers operating between 1 to 12 THz. These instruments will fill the science gap between ground-based astronomy (up to 1 THz) and mid infrared observations by James Webb Space Telescope [1]. To realize imaging spectrometers at FIR band at reasonable observation times, highly sensitive detectors with noise equivalent powers (NEP) in the order of $10^{-20}$ W/$\sqrt{\text{Hz}}$ are required. Microwave Kinetic Inductance Detectors (MKIDs) are a promising detector technology which satisfy this requirement [2] while providing multiplexing capabilities to realize large imaging arrays [3].

The state-of-the-art NEP performance of lens antenna coupled MKIDs are demonstrated previously [2]. However, classic antenna concepts borrowed from millimetre wave applications are faced with major challenges in terms of fabrication, assembly, and alignment at frequency bands above 5 THz. As a result, the community is in need for reliable and scalable technologies to bypass these challenges. Absorber based systems, although more limited in sensitivity with respect to antenna systems and unable to distinguish phase information, have been used classically at infrared and optical wavelengths as incoherent detectors. In recent years, major efforts are immerging toward designing such absorber-based systems for various applications including security and astronomical observations [4] – [6]. Currently, lens absorber based detectors are also attracting the attention of the community as a detector scheme at THz frequencies for astronomical observations [7].

The focus of this work is on the development of lens absorber coupled MKIDs, see Fig. 1, as a complementary detector scheme to the lens antenna one for FIR astronomical observations above 5 THz. Several lens absorbers coupled MKID prototypes are under fabrication and their sensitivity measurements in the dark and optical loadings are scheduled for the upcoming months.

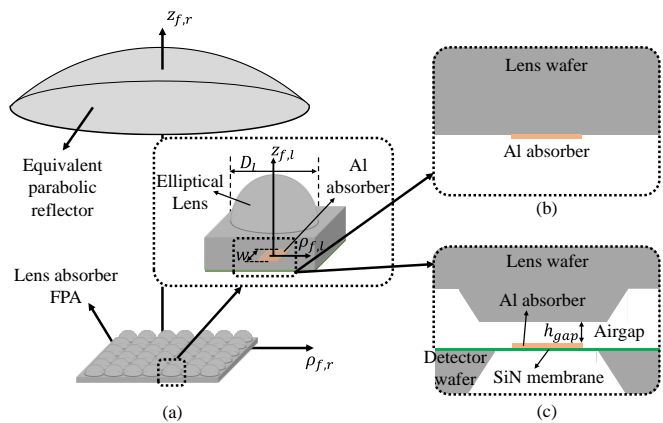

**Fig. 1.** (a) Schematic representation of a lens absorber based FPA where (b) absorber layer is fabricated over the backside of the lens wafer, and (c) absorber layer is fabricated over a SiN membrane on the detector wafer which is glued to the lens wafer.

## II. DESIGN METHODOLOGY

The considered lens absorber geometry is based on aluminum (Al) strips placed below high resistivity silicon elliptical lens elements. A standard quarter wavelength matching layer of Parylene C covers the lens elements. Two sets of detectors operating at central frequencies of 7.8 and 12 THz are considered in this work. An efficient absorption mechanism can be realized by matching the impedance of a periodic Al strip to the impedance of the incident THz radiation inside a lens element. However, strip absorber design guidelines such as [8] require narrow Al lines to reduce their inductance and provide the required impedance match at THz frequencies. On the other hand, wider lines are desired to provide better power handling for the readout of the MKIDs to improve their sensitivity. As a compromise, meander strip geometries similar to ones discussed in [9] are designed to compensate the high inductance of wide Al lines by adding capacitance effect in the unit cells.

Here, two types of stratifications are considered: I) absorber over solid substrate; and II) absorber over a thin SiN based membrane to investigate the expected enhancement of their NEP [10]. The latter stratification requires a thin airgap between the two wafer layers for stability of the membrane. The thickness of this airgap affects the absorption performance of the detectors by modifying their impedance.

The performance of the lens absorber focal plane arrays (FPAs) below the reflector system are evaluated using the computationally efficient analysis method described in [11]. In particular, the plane wave response of the considered tightly periodic absorbing meanders are modeled using the fundamental Floquet wave modes as an admittance matrix similar to [12]; and the response of the quasi-optical chain (reflector system and elliptical lens elements) is represented as a summation of plane waves using the Coherent Fourier Optics approach [13].

## III. PRELIMINARY RESULTS

The performance of several lens absorber geometries coupled to an equivalent parabolic reflector was evaluated in terms of aperture efficiency. This study was performed for the two considered stratifications, and a range of absorber side lengths, $w$, and lens focal to diameter ratios, $f_\#^l$. The diameter of the considered lens elements is $D_l = \lambda_0 f_\#^r$ where $\lambda_0$ is the wavelength in free space at the central frequency and $f_\#^r$ is the reflector's focal length to diameter ratio. This FPA sampling leads to a maximum theoretical aperture efficiency of 50%. On the other hand, the absence of a quarter wavelength backing reflector reduces the aperture efficiency by an extra factor of 77% leading to a maximum theoretical aperture efficiency of ~38%. The aperture efficiency for several stratifications is shown in Fig. 2 at the central wavelength of $\lambda_0$. As it can be seen, in the case of the membrane stratification, an airgap thickness in the order of $\lambda_0/50$ or smaller is required to reach a performance similar to the one of the solid substrate case.

The results obtained from the above parametric study is used to design four lens absorber coupled MKID FPAs with two types of stratifications operating at central frequencies of 7.8 and 12 THz.

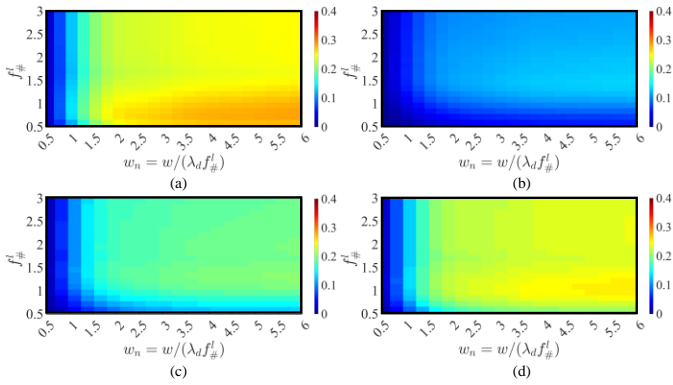

**Fig. 2.** Aperture efficiency of the considered lens absorber element coupled to an equivalent parabolic reflector as a function of the lens f-number and the absorber side length where $\lambda_d$ is the wavelength in silicon material at the central frequency: (a) solid substrate case without any airgap; absorber on SiN membrane with (b) $h_{gap} = \lambda_0/10$, (c) $h_{gap} = \lambda_0/20$, and (d) $h_{gap} = \lambda_0/50$.

## IV. CONCLUSION

Due to the relaxed requirements on the fabrication complexity, assembly and alignment, lens absorber coupled MKIDs are a promising candidate for highly sensitive detectors operating at frequencies above 5 THz. In this work, several focal plane arrays of such detectors are being developed using meandering Al strips coupled to silicon elliptical lens arrays to assess their sensitivity. The design of these geometries is achieved by resorting to a Floquet wave representation coupled to Coherent Fourier Optics. The fabrication of four FPAs is on-going and the measurement of their sensitivity is scheduled for the upcoming months.